\documentstyle[12pt,graphicx]{article}
\title{
Trapped surfaces in a hadronic fluid
 }
\author{
Neven Bili\'c and Dijana Toli\'c\\
Rudjer Bo\v{s}kovi\'{c} Institute, \\
P.O.\ Box 180, 10001 Zagreb, Croatia \\
E-mail: bilic@thphys.irb.hr, dijana.tolic@irb.hr
}
\date{\today}
\begin{document}
\maketitle
\begin{abstract}
 Pion  propagation
 in a hadronic fluid
with a nonhomogeneous relativistic flow
 is studied in terms of
 the linear sigma model.
The wave equation turns out to be equivalent to
the equation of motion for a massless
scalar field propagating in a
curved spacetime geometry.
The metric tensor
depends locally on the
soft pion dispersion relation
and the four-velocity
 of the fluid.
 For a relativistic  flow
 in curved spacetime
 the apparent and trapping horizons
 may be defined 
in the same way as in general relativity.
 An expression for
 the  analog surface gravity
 is derived.
\end{abstract}

%

\section{Introduction}

\label{introduction}

In current understanding the matter created in heavy ion collisions
behaves as a nearly perfect expanding fluid
\cite{heinz} under 
extreme conditions of  very high density
and temperature. 
This hydrodynamic behavior was observed at Brookhaven's
 Relativistic Heavy Ion Collider (RHIC) and recently
confirmed by the ALICE collaboration in Pb-Pb collisions at 
the LHC \cite {floris,cern}. 
In high energy collisions the  produced particles are predominantly pions.
The agreement between the pion production results reported in
\cite {floris}  
and the theoretical hydrodynamical model  predictions 
\cite{shen} is truly remarkable.
A realistic hydrodynamic model  
may be constructed  \cite{dumitru-kolb}
in which a transverse expansion is
superimposed on a longitudinal boost invariant expansion \cite{bjorken}.

It is often stated by particle physicists that heavy ion collisions create  mini big bangs\footnote{A few recent quotations from the press: ``What we're
 doing is reproducing the conditions that existed at the very early Universe,
 a few millionths of a second after the big bang" \cite{tuts}.
``The Large Hadron Collider has successfully created a 'mini-big bang' by smashing together lead ions instead of protons'' \cite{bbc}.
``The collisions generated mini big bangs and the highest temperatures and densities ever achieved in an experiment'' 
\cite{evans1}.}
-- events
  in which matter is created
under extreme conditions of high density and high temperature  resembling the conditions
in the early Universe a fraction of a second after the big bang. The expansion of hadronic matter
that takes place immediately after a heavy ion collision   has certain similarity with
the cosmological expansion.  However, the analogy is rather superficial since
in a cosmological expansion of spacetime after the big bang the gravity
plays the essential role, whereas high energy  collisions and subsequent expansion does not involve gravity at all.
Although the Minkowski spacetime with expanding hadronic matter can be mapped into an expanding spacetime, the 
resulting spacetime is still flat.
However,  we will demonstrate here that in
high energy collisions a much closer analogy with cosmology may be drawn owing to
the effective analog gravity with essentially curved geometry.
Various aspects of analog gravity
(for a review and extensive list of references see \cite{barcelo})
 have been studied in acoustics
\cite{visser}
optics \cite{philbin},
superfluidity
\cite{jacobson}, black hole accretion \cite{moncrief,abraham},
and hadronic fluid
near the QCD chiral phase transition
\cite{tolic}.
In this paper we study in detail
the framework of analog gravity provided by a 
hadronic fluid at nonzero temperature for the whole range of temperatures below
the chiral phase transition. We show that the analog cosmological spacetime
corresponds to a contracting Friedmann-Robertson-Walker (FRW) universe with a nontrivial apparent horizon.


Strongly interacting matter is described at the fundamental level by
a non-Abelian gauge theory called
 quantum chromodynamics (QCD).  At large distances 
or small momenta, QCD exhibits 
the phenomena of quark confinement
and chiral symmetry breaking.
At low energies, the QCD vacuum is characterized by 
a nonvanishing expectation value
\cite{shifman}:
  $\langle \bar\psi\psi\rangle \approx$ (235 
MeV)$^3$,
the so-called quark condensate,
which describes the density of quark-antiquark pairs
found in the QCD vacuum and its nonvanishing value is the manifestation of chiral symmetry breaking.
The phenomenological importance
of the chiral transition and
possible experimental signatures
have been discussed by Harris and M\"uller
\cite{harris}.

The chiral symmetry breaking and restoration at finite temperature may be 
conveniently studied using the linear sigma model \cite{bilic,bilic1}
originally proposed as a  model for
strong nuclear interactions
\cite{gell}.
Today, the linear sigma model serves as an effective
model for the low energy (low temperature)
phase of QCD. The basic model involves  
four scalar fields (three pions and a sigma meson)
and  two-flavor constituent quarks.
In the chirally symmetric phase at temperatures above the chiral transition point
the mesons are massive with equal masses and quarks are massless.
In the chirally broken phase  the pions are massless, whereas the 
quarks and sigma meson acquire a nonzero mass proportional to the
chiral condensate.
At temperatures below
the chiral phase transition point
the pions, although they are
massless, propagate slower than light
\cite{pisarski2,son1,son2} with a velocity approaching zero at the
critical temperature.
Hence, it is very likely that there exists a region where the flow velocity exceeds the pion velocity
and the analog trapped region may form.
In our previous paper \cite{tolic} we demonstrated 
that a region containing analog trapped surfaces 
forms  near the chiral phase transition.
The purpose of this paper is to study  general conditions for the formation 
of a trapped region with
the  inner boundary   as a marginally trapped surface 
which we  refer to  as the {\em analog apparent horizon}.
Our approach is based on
 the linear sigma model combined with
 a  boost invariant Bjorken type spherical expansion.
 A similar model has been previously studied in the context of disoriented
chiral condensate \cite{lampert}.

The remainder of the paper is organized as follows. 
In Sec.~\ref{chiral} we describe the properties and the dynamics 
of the chiral fluid at finite temperature.
The analog geometry of the expanding chiral fluid is studied in Sec.~\ref{analog}  
 in which we derive the condition  for  the analog apparent horizon and study
the analog Hawking effect.
 In the concluding section,
Sec.~\ref{conclusion}, we summarize our results and
discuss  physical consequences.
Finally, in the Appendix \ref{trapped} we  outline  basic notions 
related to trapped surfaces in black hole physics and cosmology.

\section{Chiral fluid}  
\label{chiral}
In this section we focus on the physics of hadrons at finite temperature  and study the properties and the dynamics 
of an expanding chiral fluid. We base our study on a linear sigma model with no fermions
which we describe in Sec.\ \ref{linear}.
In Sec. \ref{velocity} we calculate the effective velocity of pions propagating in a chiral medium.
We model the fluid expansion on a boost invariant spherical expansion of the Bjorken type
which we describe in  Sec.\ \ref{bjorken} 

\subsection{Linear sigma model}
\label{linear}
Consider a linear sigma model at finite temperature
in a general curved spacetime background.
For our purpose it is sufficient to study the model with no 
constituent fermions.
The thermal bath provides a medium which may
have an inhomogeneous velocity field.
The dynamics of mesons in  such a medium is described by
an effective chirally symmetric Lagrangian of the form
\cite{bilic2}
\begin{equation}
\label{eq1}
{\cal{L}} =
 \frac{1}{2}(a\, g^{\mu\nu}
 +b\, u^{\mu}u^{\nu})\partial_{\mu} \varphi
 \partial_{\nu} \varphi
 - \frac{m_0^2}{2}
 \varphi^2
- \frac{\lambda}{4}
 (\varphi^2)^2 ,
\end{equation}
where $u_{\mu}$ is the velocity of the fluid,
and $g_{\mu\nu}$ is the background metric.
 The mesons $\varphi\equiv (\sigma ,$
{\boldmath$\pi$})  constitute
the $(\frac{1}{2},\frac{1}{2})$
 representation of the chiral SU(2)$\times$SU(2).
The parameters
 $a$ and $b$ depend  on the local temperature $T$
 and on the parameters of the model $m_0$ and $\lambda$
 and may be calculated
in perturbation theory.
At zero temperature the medium is  absent in which case $a=1$ and $b=0$.

\begin{figure}[t]
\begin{center}
\includegraphics[width=0.6\textwidth,trim= 0 0cm 0 0cm]{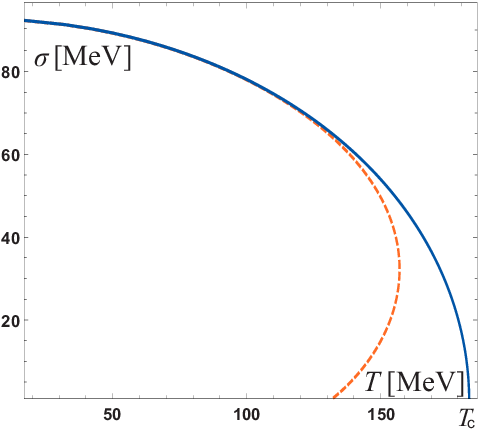}
\caption{Chiral condensate as a function of temperature
for a temperature independent (full line) and temperature dependent $m_\sigma$
(dashed line), representing a second order phase transition
and  first order (discontinuous) phase transition, respectively.
The critical temperature of the second order phase transition is indicated by 
 $T_{\rm c}$.}
 \label{fig1}
\end{center}
\end{figure}

 If $m_0^{2} < 0$ the chiral
symmetry will be spontaneously broken.
At the classical level, the $\sigma$ and $\pi$ fields develop
nonvanishing expectation values such that
at zero temperature
\begin{equation}\label{eq2}
\langle \sigma \rangle^{2} + \langle \mbox{\boldmath$\pi$} \rangle^{2}=
- \frac{m_0^{2}}{\lambda} \equiv f_{\pi}^{2} .
\end{equation}
It is convenient to choose here 
\begin{equation}\label{eq3}
\langle \pi_{i} \rangle = 0, \;\;\;\;\; \; \langle \sigma \rangle =
f_{\pi} .
\end{equation}
At nonzero temperature the expectation value
$\langle \sigma \rangle$  is  temperature dependent
and vanishes at the chiral transition point.
 Redefining the fields
\begin{equation}\label{eq9}
\varphi \rightarrow
\varphi +\varphi'(x) =
(\sigma,\mbox{\boldmath$\pi$})+
 (\sigma'(x),\mbox{\boldmath$\pi$}'(x)) ,
\end{equation}
where {\boldmath$\pi'$}
and $\sigma'$
are quantum fluctuations around the
constant values {\boldmath$\pi$} = 0 and $\sigma
=\langle \sigma \rangle$,
respectively,
we obtain
 the effective Lagrangian
 in which
the chiral symmetry is explicitly broken:
\begin{equation}\label{eq5}
{\cal{L}'} =
 \frac{1}{2}(a\, g^{\mu\nu}
 +b\, u^{\mu}u^{\nu})\partial_{\mu} \varphi'
 \partial_{\nu} \varphi'
- \frac{m_{\sigma}^{2}}{2} \sigma'^{2}
- \frac{m_{\pi}^{2}}{2}
\mbox{\boldmath$\pi$}'^{2}
-g \sigma'\varphi'^2 
- \frac{\lambda}{4}
 (\varphi'^2)^2 .
\end{equation}
The fields $\sigma'$ and $\mbox{\boldmath$\pi$}'$ correspond to the physical sigma meson and pions, respectively.
 The effective masses and
  the trilinear coupling $g$  are functions
of $\sigma$  defined as
\begin{eqnarray}\label{eq11}
m_{\sigma}^{2} & = &
 m_0^{2} + 3 \lambda \sigma^2   ,
  \nonumber \\
m_{\pi}^{2} & = & m_0^{2}+\lambda \sigma^{2}  ,
   \\
g & = & \lambda \sigma   .
  \nonumber
\end{eqnarray}
For temperatures below the
chiral transition point the meson masses are given by
\begin{equation}
m_{\pi}^2 =  0\, ; \;\;\;\;\;\;
m_{\sigma}^2 = 2\lambda \sigma^{2}  ,
\label{eq43}
\end{equation}
in agreement with the Goldstone theorem.
The temperature dependence of the chiral condensate $\sigma$
is obtained by
minimizing the thermodynamical potential
$\Omega=-(T/V) \ln Z$
with respect to
 $\sigma$
at fixed inverse temperature $\beta$.
At one loop order,
the extremum condition
reads
\cite{bilic1}
\begin{equation}
\sigma^{2}=
f_{\pi}^{2}
- 3  \int  \frac{d^3p}{(2\pi)^3}
\: \frac{1}{\omega_{\sigma}} \: n_{B} (\omega_{\sigma})
- 3  \int  \frac{d^3p}{(2\pi)^3}
\: \frac{1}{\omega_{\pi}} \: n_{B} (\omega_{\pi}) \, ,
\label{eq032}
\end{equation}
where
\begin{equation}\label{eq28}
\omega_{\pi}
=|\mbox{\boldmath $p$}|
\, ; \;\;\;\;\;\;
\omega_{\sigma}
=(\mbox{\boldmath $p$}^2+m_{\sigma}^{2})^{1/2}
\end{equation}
 are the energies of the $\pi$ and
 $\sigma$ particles, respectively,
and
\begin{equation}\label{eq30}
n_{B}(\omega) = \frac{1}{e^{\beta \omega} - 1}
\end{equation}
is the Bose-Einstein distribution function.
Equation (\ref{eq032}) has been derived from the
zero order thermodynamical  potential
with  meson masses at one loop order
\cite{bilic1}. This approximation corresponds to the
leading order in 
$1/N$ expansion, where $N$ is the
number of scalar fields \cite{meyer}.
In our case, $N=4$.
The right-hand side of (\ref{eq032}) depends on
$\sigma$ through the mass
$m_{\sigma}$ given by (\ref{eq43}).
The behavior of $\sigma$ near the critical
   temperature should be analyzed with special care.
 A straightforward solution to  (\ref{eq032})
 as a function of temperature exhibits a weak first order
 phase transition \cite{bilic1,rod}.
 However,  Pisarski and Wilczek have shown on general grounds
  that  the phase transition in SU(2)$\times$SU(2) chiral models should be of second order \cite{pis}.
  Hence, it is generally believed that a  first order
   transition is an artifact of 
   the one loop approximation.
      Two loop calculations \cite{baacke} make an improvement and 
   confirm the general analysis of  \cite{pis}.
 It is possible to  mimic the second order phase  transition
even with (\ref{eq032}) 
 by   making the $\sigma$-meson  mass  
temperature independent all the way up to the critical temperature
and equal to its zero temperature mean field  value given by
\begin{equation}
m_{\sigma}^2 = 2\lambda f_\pi^2 ,
\label{eq31}
\end{equation}
instead of (\ref{eq43}). We fix the coupling $\lambda$ from 
the values of $m_\sigma$ and $f_\pi$ for which we take $m_\sigma=1$  GeV 
and $f_\pi = 92.4$ MeV
 as a 
phenomenological input.
In Fig.\ \ref{fig1} we plot the solutions to  (\ref{eq032}) for
both temperature dependent and temperature independent $m_\sigma$ exhibiting apparent
first and second order phase transitions, respectively.
In the rest of our paper we employ the solution that corresponds to the second order phase transition.
For our choice of parameters we find numerically $T_{\rm c}=182.822$ MeV.

The propagation of pions is governed by the equation of motion
\begin{equation}
\frac{1}{\sqrt{-g}}
\partial_{\mu}
\left[
{\sqrt{-g}}\,
( a\, g^{\mu\nu}+ b\,
u^{\mu}
u^{\nu}) \partial_{\nu}\mbox{\boldmath{$\pi$}}\right]
+V(\sigma,
\mbox{\boldmath{$\pi$}})
\mbox{\boldmath{$\pi$}}=0,
\label{eq013}
\end{equation}
where 
\begin{equation}
V(\sigma,
\mbox{\boldmath{$\pi$}})=
m_\pi^2 + g\sigma+ \lambda (\sigma^2+\mbox{\boldmath{$\pi$}}^2)
\label{eq213}
\end{equation}
 is the interaction potential.
In the comoving reference frame
  in flat spacetime,
  Eq. (\ref{eq013}) reduces to
  the  wave equation
\begin{equation}
(\partial_t^2 -
c_{\pi}^2
\Delta +\frac{c_{\pi}^2}{a}V)
\mbox{\boldmath{$\pi$}}=0 ,
\label{eq014}
\end{equation}
where the quantity $c_{\pi}$,
defined  by
\begin{equation}
c_{\pi}^2=\left(1+\frac{b}{a}\right)^{-1} ,
\label{eq015}
\end{equation}
is the pion velocity. As we shall demonstrate in the next section, the constants $a$ and $b$
may be derived from the finite-temperature perturbation
expansion of the pion self energy.


\subsection{Pion velocity}
\label{velocity}
At temperatures below
the chiral transition point,
the pions are massless.
However, the velocity of massless particles in a medium is not necessarily equal to the velocity of light —
in the chiral fluid pions usually propagate slower than light.\footnote{If the chiral fermions are present,
pions become superluminal
in certain range of temperature and baryon chemical
potential \cite{bilic2}.}
The pion velocity in a sigma model at finite temperature
has been calculated at one loop level by Pisarski and Tytgat
in the low temperature approximation
\cite{pisarski2} and by Son and Stephanov for temperatures
close the chiral transition point \cite{son1,son2}.
It has been found that the pion velocity vanishes
as one approaches the critical temperature.
Here we summarize the calculation of the parameters $a$ and $b$  in the
entire range of temperatures in the
chiral symmetry broken phase \cite{bilic2}.

The pion velocity may be derived from the
 self-energy  $\Sigma(q,T)$
in the limit where the external momentum
$q$ approaches 0.
For a flat background geometry $g_{\mu\nu}=\eta_{\mu\nu}$,
the inverse pion propagator $\Delta^{-1}$, derived directly from
the effective Lagrangian (\ref{eq5}) as
\begin{equation}
 \Delta^{-1}=a q^{\mu}q_{\mu}
 +b(q^{\mu}u_{\mu})^2 -m_{\pi}^2 ,
  \label{eq200}
\end{equation}
may in the limit $q\rightarrow 0$ be expressed in the form
\begin{equation}
 Z_{\pi}\Delta^{-1}=
  q^{\mu}q_{\mu}-
 \frac{1}{2!}
 q^{\alpha}q^{\beta}\left[\frac{\partial}{\partial q^{\alpha}}
 \frac{\partial}{\partial q^{\beta}}
 (\Sigma(q,T)
-  \Sigma(q,0)) \right]_{q=0}
+\dots ,
  \label{eq201}
\end{equation}
where the ellipsis denotes the terms of higher order in
$q^{\mu}$.
The $q^{\mu}$ independent term of the self-energy
absorbs in the renormalized pion mass,
equal to zero in the chiral symmetry broken phase.
The subtracted $T=0$ term  has been
absorbed in the wave function renormalization factor $Z_{\pi}$.
By comparing this equation with
Eq.\ (\ref{eq200}) 
written in the comoving frame
 as
\begin{equation}
 \Delta^{-1}=(a+b)q_0^2
 -a \mbox{\boldmath $q$}^2-m_\pi^2,
  \label{eq202}
\end{equation}
we can express
 the parameters $a$ and $b$,
 and hence the pion
velocity, in terms of second derivatives of
$\Sigma(q,T)$ evaluated at $q^{\mu}=0$.
At one loop level the only diagram that
gives a nontrivial q dependence of $\Sigma$ is the bubble
diagram. Subtracting the $T=0$ term
 one finds
\cite{son2}
\begin{eqnarray}
\Sigma(q)
  \!&\! \equiv \!&\!
\Sigma(q,T)
-  \Sigma(q,0)
  = -4{g}^2 \int\!
\frac{d^3p}{(2\pi)^3}
\frac{1}{2\omega_{\pi}
2\omega_{\sigma,q}}
\nonumber\\
  \!&\!\!&\!
  \left\{ [n_B(\omega_{\pi})+
n_B(\omega_{\sigma,q})]
\left(\frac{1}{\omega_{\sigma,q}+
\omega_{\pi}}
  + \frac{1}{\omega_{\sigma,q}+
\omega_{\pi}+q_0}\right)\right. \nonumber\\
  \!&\!\!&\!
   + \left. [n_B(\omega_{\pi})-
n_B(\omega_{\sigma,q})]\left(
  \frac{1}{\omega_{\sigma,q}-
\omega_{\pi}} +
  \frac{1}{\omega_{\sigma,q}-
\omega_{\pi}+ q_0}\right)\right\} ,
  \label{eq203}
\end{eqnarray}
where
$\omega_{\sigma,q}=
[(\mbox{\boldmath $p$}-
                \mbox{\boldmath $q$})^2
+m_\sigma^2]^{1/2}$.
Here we take $m_\sigma$ to be a function of $\sigma$ through
Eq. (\ref{eq43}).
A straightforward evaluation of the second derivatives of
$\Sigma(q)$ at $q_{\mu}=0$ yields
\begin{equation}
a =
  1+  \frac{16 {g}^2}{m_{\sigma}^4} \int\!
\frac{d^3p}{(2\pi)^3}
  \left[ \frac{n_B(\omega_{\pi})}{4\omega_{\pi}}+
\frac{n_B(\omega_{\sigma})
}{4\omega_{\sigma}}
 - \frac{1}{3}
      \frac{\omega_{\pi}^2}{m_{\sigma}^2}
   \left(
   \frac{n_B(\omega_{\pi})}{\omega_{\pi}} -
\frac{n_B(\omega_{\sigma})
}{\omega_{\sigma}}
\right)\right]  ,
  \label{eq204}
\end{equation}
\begin{equation}
b =
   \frac{16{g}^2}{m_{\sigma}^4} \int\!
\frac{d^3p}{(2\pi)^3}
  \left[
  \frac{\omega_{\pi} n_B(\omega_{\pi})
  }{m_{\sigma}^2}-
\frac{\omega_{\sigma}
n_B(\omega_{\sigma})
  }{m_{\sigma}^2}
 + \frac{1}{3}
      \frac{\omega_{\pi}^2}{m_{\sigma}^2}
   \left(
   \frac{n_B(\omega_{\pi})}{\omega_{\pi}} -
\frac{n_B(\omega_{\sigma})
}{\omega_{\sigma}}
\right)\right]  .
  \label{eq205}
\end{equation}
The pion velocity $c_{\pi}$ as given by  (\ref{eq015})
depends on temperature explicitly 
through the thermal distribution function $n_B$
and implicitly through  the
chiral condensate $\sigma$ given by Eq.\ (\ref{eq032}).
\begin{figure}[t]
\begin{center}
\includegraphics[width=0.6\textwidth,trim= 0 0cm 0 0cm]{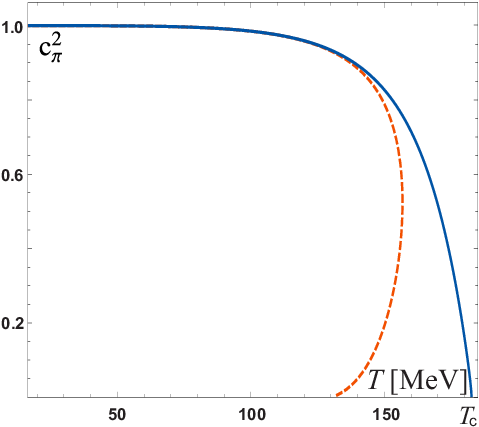}
\caption{
Pion velocity as a function of temperature 
for a temperature independent (full line) and temperature dependent $m_\sigma$
(dashed line). The critical temperature of the second order phase transition is indicated by 
 $T_{\rm c}$.}
 \label{fig1a}
\end{center}
\end{figure}
In Fig.\ \ref{fig1a} we plot $c_{\pi}$ as a function of temperature
corresponding to two solutions depicted in the left panel.

\subsection{Spherical Bjorken expansion}
\label{bjorken}
In order to explore the analogy between the chiral fluid and cosmological expansions,
we consider a boost invariant 
spherically symmetric  Bjorken type expansion \cite{bjorken}
in Minkowski background spacetime.
In radial coordinates
$x^\mu=(t,r,\vartheta,\varphi)$, the fluid velocity is  given by
\begin{equation}
u^\mu=(\gamma,\gamma v, 0,0)= (t/\tau, r/\tau,0,0),
\label{eq144}
\end{equation}
where $v=r/t $ is the radial three-velocity  and $\tau=\sqrt{t^2-r^2}$ is the 
{\em proper time}.
Using the so-called {\em radial rapidity}
\begin{equation}
y=\frac{1}{2} \ln \frac{t+r}{t-r} ,
\label{eq145}
\end{equation}
the velocity is expressed as
\begin{equation}
u^\mu=(\cosh y,\sinh y,0, 0),
\label{eq146}
\end{equation}
and hence, the radial three-velocity is
\begin{equation}
v=\tanh y.
\label{eq246}
\end{equation}
It is convenient to change  $(t,r,\vartheta,\varphi)$ to new coordinates
$(\tau,y,\vartheta,\varphi)$ via the transformation 
\begin{eqnarray}
& &t=\tau \cosh y ,
\nonumber \\
& & r=\tau \sinh y .
\label{eq147}
\end{eqnarray}
In these coordinates the background Minkowski metric takes the form
\begin{equation}
g_{\mu\nu}
 =
\left(\begin{array}{cccc}
1          &          &  &   \\
           & -\tau^2  &     &     \\
           &          & -\tau^2\sinh^2\! y &   \\
           &          &                  & -\tau^2\sinh^2\! y \sin^2 \vartheta  
\end{array} \right).
\label{eq218}
\end{equation}
and the velocity components become  $u^\mu=(1,0,0,0)$.
Hence,
the new coordinate frame is comoving.
The metric (\ref{eq218}) corresponds to the Milne cosmological model — a 
homogeneous, isotropic, expanding universe
with the cosmological scale  $a=\tau$ and negative spatial curvature.

The functional dependence of $T$ on $\tau$ follows from
 the energy-momentum conservation.
For a perfect relativistic fluid
the energy-momentum tensor is
given by
\begin{equation}
T_{\mu\nu}=(p+\rho) u_{\mu}u_{\nu}-p g_{\mu\nu}   ,
\label{eq001}
\end{equation}
where $p$ and $\rho$ denote, respectively, the  pressure and 
the energy density of the fluid. From the energy-momentum conservation
\begin{equation}
{T^{\mu\nu}}_{;\nu}=0
\label{eq102}
\end{equation}
 applied to  (\ref{eq001})
we find
\begin{equation}
u^\mu \rho_{,\mu}+(p+\rho){u^\mu}_{;\mu}=0,
\label{eq003}
\end{equation}
where the subscript $;\mu$ denotes the covariant derivative  associated with
the background metric.
Since our fluid is dominated by massles pions at nonzero temperature, it is a reasonable approximation
to assume the equation of state $p=\rho/3$
of an ideal gas of massless bosons.
Then, Eq.\ (\ref{eq003}) in comoving coordinates reads
\begin{equation}
\frac{\partial\rho}{\partial\tau}  + \frac{4\rho}{\tau}  =0
\label{eq148}
\end{equation}
with the solution
\begin{equation}
\rho=\rho_0 \left(\frac{\tau_0}{\tau}\right)^4.
\label{eq006}
\end{equation}
This expression combined  with the density of the pion gas 
\cite{landau}
\begin{equation}
\rho=\frac{\pi^2}{10}T^4,
\label{eq106}
\end{equation}
implies the temperature profile
\begin{equation}
T=T_0 \frac{\tau_0}{\tau}.
\label{eq007}
\end{equation}
The constants $T_0$ and $\tau_0$ may be fixed from 
the phenomenology of high energy collisions. For example, if we choose $T_0=1 {\rm GeV}$,
then a typical value of
$\rho=1 {\rm GeV/fm^3}$ at  $\tau\approx 5 \,{\rm fm}$ 
\cite{kolb-russkikh}
is obtained with  $\tau_0 = 1.5\, {\rm fm}$. In our case, with these values the interesting range of temperatures $T$
between 100 and 200 MeV corresponds to $\tau$ between 15 and 30 fm. In the following we work with $T_0=1{\rm GeV}$ and
keep $\tau_0$ unspecified so that physical quantities of dimension of time or length are expressed in units
of $\tau_0$.

\section{Analog cosmology}
\label{analog}
In this section we turn to study the analog metric and formation and properties of  
the  apparent horizon in an expanding chiral fluid.
To this end  we outline the formalism in the first subsection
and  derive a condition for the apparent horizon for a general hyperbolic spacetime.
In Sec. \ref{horizon} we derive the analog metric 
for the expanding chiral fluid and study
the properties of the analog apparent horizon. 
Then, in Sec. \ref{surface}  we exploit the Kodama-Hayward definition of surface gravity
to derive the Hawking temperature as a function of the  parameters of the chiral fluid, 
in particular, as a function of the local fluid temperature.

\subsection{Radial null geodesics}
\label{radial}
To study the  apparent horizon in an expanding chiral fluid 
we need to examine the behavior of radial null geodesics of the analog metric
which we shall derive in Sec. \ref{horizon}.
With hindsight, we first consider a spacetime of the form
\begin{equation}
ds^2= \beta(\tau)^2 d\tau^2 -\alpha(\tau)^2 (dy^2 + \sinh^2\! y \, d\Omega^2),
\label{eq008}
\end{equation}
where $\beta$ and $\alpha$ are arbitrary functions of $\tau$. 
The metric tensor is
\begin{equation}
G_{\mu\nu}
 =
\left(\begin{array}{cccc}
  \beta^2  &          &  &   \\
           & -\alpha^2  &     &     \\
           &          & -\alpha^2\sinh^2\! y &   \\
           &          &                  & -\alpha^2\sinh^2\! y \sin^2 \vartheta 
\end{array} \right).
\label{eq243}
\end{equation}
This metric represents the class of hyperbolic ($k=-1$) FRW spacetimes
including the flat spacetime example (\ref{eq218}).
We denote by 
$l_+^\mu$ and $l_-^\mu$ the vectors tangent to outgoing and ingoing 
affinely parametrized 
radial null geodesics 
normal to a spherical two-dimensional surface $S$. The tangent vectors are null with respect to the metric
(\ref{eq243}), i.e.,
\begin{equation}
G_{\mu\nu}l_+^\mu l_+^\nu =G_{\mu\nu}l_-^\mu l_-^\nu= 0 .
\label{eq143}
\end{equation}
Using the geodesic equation
\begin{equation}
l^\mu \nabla_\mu{l^\nu}=0,
\label{eq009}
\end{equation}
where the symbol $\nabla_\mu$  denotes the covariant derivative associated with
the metric (\ref{eq243}),
one easily finds  the tangent null vectors  corresponding to four types of radial null geodesics,
\begin{equation}
l_\pm^\mu  =q_\pm \alpha^{-1}\left(  \beta^{-1}, \pm\alpha^{-1},0,0\right),
\label{eq010}
\end{equation}
tangent to future directed and
\begin{equation}
l_\pm^\mu  =\tilde{q}_\pm \alpha^{-1}\left( - \beta^{-1}, \pm\alpha^{-1},0,0\right),
\label{eq011}
\end{equation}
to past directed null geodesics , where $q_+$, $q_-$, $\tilde{q}_+$, and $\tilde{q}_-$ 
are  arbitrary positive constants.
The corresponding affine parameters  $\lambda_+$ and $\lambda_-$ for the outgoing and ingoing null geodesics,
respectively, are found to satisfy
\begin{equation}
\frac{d\lambda_\pm}{d\tau}= \frac{1}{q_\pm}\alpha\beta 
\label{eq012}
\end{equation}
 for future directed and 
\begin{equation}
\frac{d\lambda_\pm}{d\tau}=- \frac{1}{\tilde{q}_\pm}\alpha\beta ,
\label{eq019}
\end{equation}
for past directed null geodesics.
For simplicity, from now on  we set $q_+ =q_-=\tilde{q}_+=\tilde{q}_-=1$.

The null vectors
 $l_+^\mu$ and $l_-^\mu$ point towards increasing and decreasing $y$, respectively.
Hence, we adopt the usual convention and refer 
to $l_+^\mu$ ($l_-^\mu$) and the corresponding null geodesics as outgoing (ingoing)
although increasing (decreasing)  $y$ does not necessarily imply increasing (decreasing) of the radial coordinate $r$.
As we move along a geodesic the changes of the coordinates $\tau$ and  $y$ are
subject to the condition
$ds=0$ of radial null geodesics,
i.e.,  
\begin{equation}
d\tau=\pm \frac{\alpha}{\beta} dy 
\label{eq017}
\end{equation}
along the geodesic. Here the signs
determine whether $y$ is increasing or decreasing as we move along the geodesic.
For example, for  future directed null geodesics,
it follows from  (\ref{eq012}) and (\ref{eq017}) that an outgoing geodesic is directed along increasing $y$, i.e., $y$ increases with $\lambda_+$, whereas an ingoing geodesic is directed along decreasing $y$, i.e., $y$ 
decreases with $\lambda_-$.

The key roles in the study of trapped surfaces are played by the 
expansion parameters  $\varepsilon_+$ and $\varepsilon_-$,
\begin{equation}
\varepsilon_\pm=\nabla_\mu l_\pm^\mu  
\label{eq244}
\end{equation}
of outgoing and ingoing null geodesics,
respectively.
Particularly important are the values of  $\varepsilon_+$ and $\varepsilon_-$
and their Lie derivatives, 
\begin{equation}
\frac{d\varepsilon_+}{d\lambda_-} \equiv
l^\mu_-\partial_\mu \varepsilon_+ ;
\hspace{1cm}
\frac{d\varepsilon_-}{d\lambda_+} \equiv
l^\mu_+\partial_\mu \varepsilon_-
\end{equation}
in the neighborhood of a marginally trapped surface.
As we shall shortly demonstrate, the relevant marginally trapped surface in the expanding chiral fluid
is future inner marginally trapped.
According to our convention described in the Appendix \ref{trapped},
a two-dimensional surface $H$ is said to be {\em future inner marginally trapped} 
if the future directed null expansions 
 on $H$ satisfy the conditions:  $\varepsilon_+|_H=0$, $l_-^\mu\partial_\mu\varepsilon_+|_H>0$ 
 and $\varepsilon_-|_H<0$.
The future inner  marginally trapped surface is the {\em inner} boundary of a future trapped region
consisting of  trapped surfaces with negative ingoing and outgoing null expansions.
 From now on we refer to this surface  as
the {\em apparent horizon}.

From (\ref{eq010}) and (\ref{eq011})  we find
\begin{equation}
\varepsilon_\pm  =\frac{2}{\alpha^2}\left(\frac{\dot{\alpha}}{\beta}\pm \frac{1}{\tanh y} \right)
\label{eq110}
\end{equation}
for future directed 
and 
\begin{equation}
\varepsilon_\pm  =\frac{2}{\alpha^2}\left(-\frac{\dot{\alpha}}{\beta}\pm \frac{1}{\tanh y} \right)
\label{eq111}
\end{equation}
 for  past directed radial null geodesics,
 where the overdot denotes a partial derivative with respect to $\tau$. 
The respective Lie derivatives are given by
\begin{equation}
\frac{d\varepsilon_\pm}{d\lambda_\mp} \equiv
l^\mu_\mp\partial_\mu \varepsilon_\pm
  =
\frac{2}{\alpha^2\beta^2}\left[\frac{\ddot{\alpha}}{\alpha}
-\frac{\dot{\alpha}\dot{\beta}}{\alpha\beta}
-\left( 1-\frac{1}{\tanh^2\! y} \right)\frac{\beta^2}{\alpha^2}\right] 
-\frac{2\dot{\alpha}}{\alpha^2\beta}\varepsilon_\pm,
\label{eq112}
\end{equation}
for future directed 
and 
\begin{equation}
\frac{d\varepsilon_\pm}{d\lambda_\mp} \equiv
l^\mu_\mp\partial_\mu \varepsilon_\pm =
\frac{2}{\alpha^2\beta^2}\left[\frac{\ddot{\alpha}}{\alpha}
-\frac{\dot{\alpha}\dot{\beta}}{\alpha\beta}
+\left( 1-\frac{1}{\tanh^2\! y} \right)\frac{\beta^2}{\alpha^2}\right] 
+\frac{2\dot{\alpha}}{\alpha^2\beta}\varepsilon_\pm, 
\label{eq113}
\end{equation}
 for past directed radial null geodesics.

For a spherically symmetric spacetime, the condition that  one of the null 
expansions  vanishes on the apparent horizon $H$
 is equivalent to the condition that the vector $n_\mu$, normal to the surface of spherical symmetry,
is null on $H$.
  In other words, the condition
 \begin{equation}
\nabla_\mu l^\mu|_H=0, 
\label{eq115}
\end{equation}
where $l^\mu$ denotes  either $l^\mu_+$ or $l^\mu_+$, is equivalent to the condition
\begin{equation}
G^{\mu\nu}n_\mu n_\nu|_H=0.
\label{eq116}
\end{equation}
This may be seen as follows.
For the metric (\ref{eq243}) the normal $n_\mu$ is given by
\begin{equation}
n_\mu=\partial_\mu (\alpha \sinh y).
\label{eq114}
\end{equation}
The expansion $\varepsilon_+$ (or $\varepsilon_-$) defined in (\ref{eq244})
may be written as
 \begin{equation}
\nabla_\mu l^\mu=\frac{1}{\sqrt{-G}}\partial_\mu (\sqrt{-G} l^\mu) 
=\frac{1}{\sqrt{-h}}\partial_\mu (\sqrt{-h} l^\mu) + \frac{2}{\alpha\sinh y}l^\mu n_\mu 
\label{eq120}
\end{equation}
where $h$ denotes the determinant of the  metric
\begin{equation}
h_{\alpha\beta}
 =
\left(\begin{array}{cc}
\beta^2    &     0      \\
     0      & -\alpha^2      \\
\end{array} \right),
\label{eq122}
\end{equation}
of the two-dimensional space normal to the surface of spherical symmetry.
It may be shown that the first term on the right-hand side of (\ref{eq120})
vanishes identically by the geodesic equation. Hence,  the expansion $\nabla_\mu l^\mu$ 
vanishes 
on $H$ if and only if  
\begin{equation}
l^\beta n_\beta|_H=0. 
\label{eq123}
\end{equation}
Suppose one of the expansions vanishes on $H$, i.e., Eq. (\ref{eq123}) holds
for either $l^\mu_+$ or $l^\mu_-$. 
Since $l^\mu$ is null and both $l^\mu$ and $n^\mu$ are normal to $H$ and hence 
tangent to the two-dimensional space $(\tau, y)$ with the metric (\ref{eq122}),
Eq.\ (\ref{eq123}) implies  $h_{\alpha\beta}n^\alpha n^\beta|_H=0$.
Hence, $\nabla_\mu l^\mu|_H=0$ implies $G^{\mu\nu}n_\mu n_\nu|_H=0$.

To prove the reverse it is sufficient to show
that $l^\beta_+ n_\beta\neq 0$ and $l^\beta_- n_\beta\neq 0$ implies $h_{\alpha\beta}n^\alpha n^\beta \neq 0$,
which may be easily shown for a general two-dimensional metric in diagonal gauge.
Then, the following  statement holds: the vanishing of  $h_{\alpha\beta}n^\alpha n^\beta$ on $H$ implies  
 either $l^\beta_+ n_\beta|_H=0$ or $l^\beta_- n_\beta|_H=0$.
 This 
together with (\ref{eq120}),
implies  either
 $\varepsilon_+|_H=0$ or $\varepsilon_-|_H=0$.

Either from  (\ref{eq115}) or (\ref{eq116}) one finds the condition for the apparent horizon
\begin{equation}
\frac{\dot{\alpha}}{\beta}\pm \frac{1}{\tanh y} =0
\label{eq117}
\end{equation}

\subsection{Analog  horizons}
\label{horizon}
Next we derive the analog metric and define the analog Hubble and the apparent horizons.
Equation (\ref{eq013}) may be written in the form
\cite{moncrief,bilic3,visser2}
\begin{equation}
\frac{1}{\sqrt{-G}}\,
\partial_{\mu}
(\sqrt{-G}\,
G^{\mu\nu})
\partial_{\nu}
\mbox{\boldmath{$\pi$}}
+\frac{c_{\pi}^2}{a}V(\sigma,
\mbox{\boldmath{$\pi$}})
\mbox{\boldmath{$\pi$}}=0 ,
\label{eq028}
\end{equation}
with the analog metric tensor,
its inverse, and its determinant given by
\begin{equation}
G_{\mu\nu} =\frac{a}{c_{\pi}}
[g_{\mu\nu}-(1-c_{\pi}^2)u_{\mu}u_{\nu}] ,
\label{eq022}
\end{equation}
\begin{equation}
G^{\mu\nu} =
\frac{c_{\pi}}{a}
\left[g^{\mu\nu}-(1-\frac{1}{c_{\pi}^2})u^{\mu}u^{\nu}
\right],
\label{eq029}
\end{equation}
\begin{equation}
G = \frac{a^4}{c_{\pi}^2}g .
\label{eq030}
\end{equation}
Hence, the pion field propagates in a (3+1)-dimensional
effective geometry described by the metric
$G_{\mu\nu}$.

In the comoving coordinate frame defined by the coordinate transformation
(\ref{eq147})
the
velocity is $u^\mu=(1,0,0,0)$ 
and, as a consequence,
the analog metric (\ref{eq022}) is diagonal
\begin{equation}
G_{\mu\nu}
 =\frac{a}{c_\pi}
\left(\begin{array}{cccc}
c_\pi^2    &          &  &   \\
           & -\tau^2  &     &     \\
           &          & -\tau^2\sinh^2\! y &   \\
           &          &                  & -\tau^2\sinh^2\! y \sin^2 \vartheta  
\end{array} \right).
\label{eq018}
\end{equation}
Here, the parameters $a$ and $c_\pi$ are functions of the temperature $T$, which in turn
is a function of $\tau$. In the following we assume
that these functions are positive.
The metric  (\ref{eq018}) is precisely of the form
(\ref{eq243})  with
\begin{equation}
\beta(\tau)=\sqrt{ac_\pi}  ; \hspace{0.5in} \alpha(\tau)=\tau\sqrt{\frac{a}{c_\pi}} .
\label{eq108}
\end{equation}
The physical range of $\tau$ is fixed by Eq.~(\ref{eq007}) since the available temperature ranges are
between $T=0$ and $T=T_{\rm c}$. Hence, the proper time range is
$\tau_{\rm c}\leq \tau < \infty$, where the  critical value $\tau_{\rm c}$ is related to the critical temperature as
$\tau_{\rm c}/\tau_0= T_0/T_{\rm c}$. The metric is singular at $\tau=\tau_{\rm c}$.

\begin{figure}[t]
\begin{center}
\includegraphics[width=0.9\textwidth,trim= 0 0cm 0 0cm]{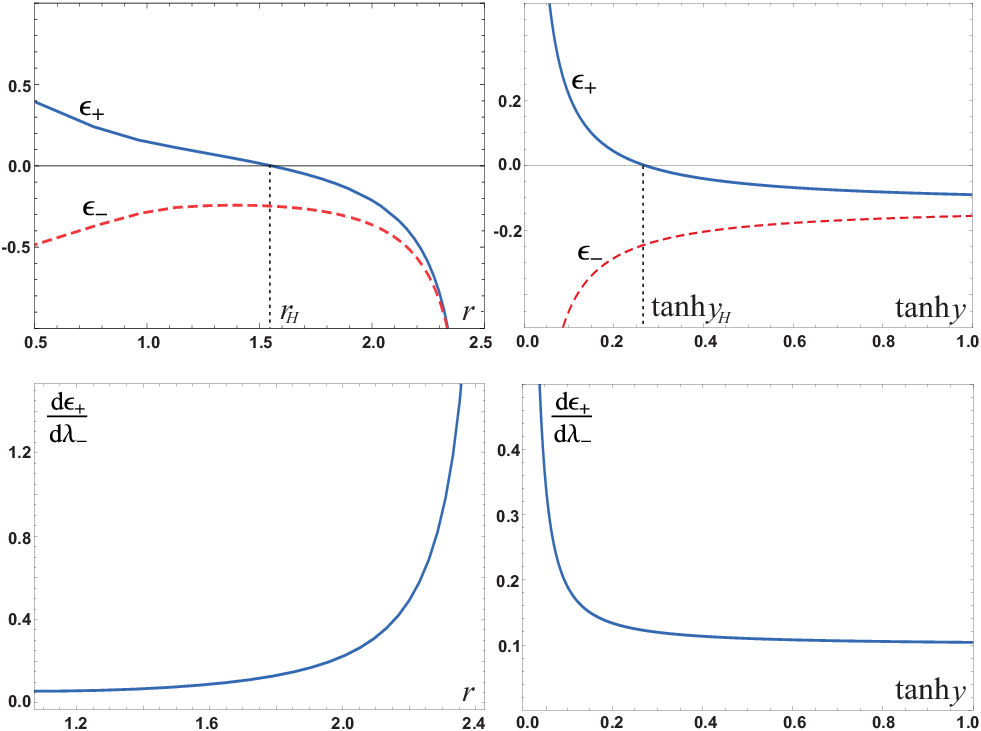}
\caption{Null expansions $\varepsilon_+$ and $\varepsilon_-$ as functions of $r$ (top left panel) and $y$ 
 (top right panel) for  fixed  $t=6\tau_0$  and fixed $\tau=5.77 \tau_0$, respectively.
Similarly, the bottom panels depict the derivative of $\varepsilon_+$ as functions of $r$ and $y$ for fixed
$t$ and $\tau$, respectively.}
\label{fig2}
\end{center}
\end{figure}

In Fig.\ \ref{fig2}  we plot the expansions $\varepsilon_+$ and $\varepsilon_-$ 
of outgoing and ingoing radial null geodesics, respectively, as functions of $r$   
  for an arbitrarily chosen fixed time $t=6 t_0$ and, similarly, as functions of $y$ for
a fixed $\tau=5.77 \tau_0$. 
In the lower two panels we plot the derivative of the outgoing null expansion $\varepsilon_+$ 
along the ingoing null geodesic.  The outgoing null expansion 
 decreases with increasing $r$ from positive to negative values and vanishes at the point $r=r_H$, 
whereas the ingoing null expansion remains negative. At this point the derivative of $\varepsilon_+$
with respect to $\lambda_-$ is positive.
According to the standard convention described
 in  Appendix \ref{trapped}
the region $\{r>r_H, t=6\tau_0\}$ is future trapped and the location 
$r_H$
 marks its  inner boundary. Thus, the sphere at $r_H$ is future inner marginally trapped.

\begin{figure}[t]
\begin{center}
\includegraphics[width=0.6\textwidth,trim= 0 0cm 0cm 0cm]{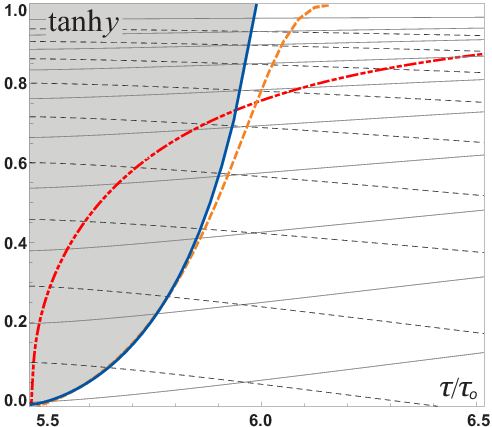}
\caption{Spacetime diagram of outgoing (full line) and ingoing (dashed line)
radial null geodesics in $(\tau,y)$ coordinates. 
The shaded area 
represents the evolution of the trapped region.
The trapping horizon is represented by the full bold line
with the endpoint at  $\tau=\tau_{\rm max}=6.0182\tau_0$.
The dashed and dash-dotted bold lines
represent the evolution of the analog and naive Hubble horizons,
respectively.}
 \label{fig3}
\end{center}
\end{figure}

Spacetime diagram corresponding to the metric (\ref{eq018})
is presented in Fig.~\ref{fig3}, showing  future directed radial null geodesics.
The origin in the plots in both panels  corresponds to the critical value $\tau_{\rm c}$
at which $c_\pi$ vanishes.
 Numerically, with the chosen  $T_0=1$ GeV we have
$\tau_{\rm c}/\tau_0=5.47$.
The geodesic lines are constructed using 
\begin{equation}
y= \pm \int_{\tau_{\rm c}}^\tau d\tau' c_\pi(\tau')/\tau' +{\rm const}
 \label{eq216}
\end{equation}
which follows from (\ref{eq017}) with (\ref{eq108}).
As mentioned in Sec.\ \ref{radial},
increasing (decreasing)  $y$ does not necessarily imply increasing (decreasing) of the radial coordinate $r$.
With the help of the coordinate transformation (\ref{eq147}) the shift  $dy$ along a geodesic 
 may be expressed in terms of $dr$
 \begin{equation}
dy= 
 \frac{c_\pi}{( c_\pi\pm v )\tau\cosh y } dr .
\label{eq016}
\end{equation}
where we have used (\ref{eq246}) and (\ref{eq108}).
We note that if $v<c_\pi$, increasing (decreasing) $y$
  corresponds to increasing (decreasing) $r$ for both signs, whereas if $v>c_\pi$ 
increasing $y$ corresponds to increasing $r$ for an outgoing
and decreasing $r$ for an  ingoing geodesic.


Using (\ref{eq216}) we introduce null coordinates
\begin{equation}
w= \frac{1}{2}\left(y+\int_{\tau_{\rm c}}^\tau d\tau' \beta(\tau')/\alpha(\tau')\right);
\hspace{1cm} 
u= \frac{1}{2}\left(-y+\int_{\tau_{\rm c}}^\tau d\tau' \beta(\tau')/\alpha(\tau')\right),
 \label{eq316}
\end{equation}
ranging in the intervals $[0,+\infty)$ and $(-\infty,+\infty)$, respectively,  with a
condition 
  $0\leq w\pm u < \infty$.
In these coordinates the metric (\ref{eq008}) becomes
\begin{equation}
ds^2= \alpha^2\left( 4 du dw - \sinh^2\! (w-u) \, d\Omega^2\right).
\label{eq208}
\end{equation}
The singularity at $\tau=\tau_{\rm c}$ is mapped onto the entire  $u+w=0$ line.
\begin{figure}[t]
\begin{center}
\includegraphics[width=0.6\textwidth,trim= 0 0cm 0 0cm]{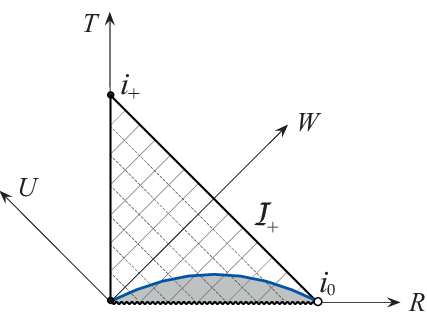}
\caption{Conformal diagram for the analog spacetime described by the metric (\ref{eq018}).
The lines  parallel to the
$W$ and $U$ axes, are future directed outgoing and ingoing null geodesics, respectively.
The singularity at $\tau=\tau_{\rm c}$ is represented by the wavy line and 
the  apparent horizon by the full line.
The shaded area in between represents the evolution of the trapped region.
The spacelike and  future timelike infinities are denoted by $i_0$ and $i_+$,
respectively. The  future null infinity ${\cal I}_+$ is represented by the line $T+R=\sqrt{2}$.}
\label{fig3a}
\end{center}
\end{figure}
Next, we compactify the spacetime using the coordinate transformation
\begin{equation}
W=\tanh w;
\hspace{1cm} 
U= \tanh u.
 \label{eq317}
\end{equation}
The coordinates $W$ and $U$ range in the intervals $[0,1]$ and $[-1,1]$, respectively,  with a
condition   $0\leq W\pm U \leq 2$.
Furthermore,  the rotation
\begin{equation}
T=\frac{1}{\sqrt{2}}(W+U);
\hspace{1cm} 
R= \frac{1}{\sqrt{2}}(W-U) ,
 \label{eq319}
\end{equation}
brings the metric to a conformally flat form
\begin{equation}
ds^2= \frac{2\alpha^2}{(1-U^2)(1-W^2)}\left[ dT^2 -dR^2 -  R^2 \, d\Omega^2\right],
\label{eq309}
\end{equation}
where both coordinates $R$ and $T$  range in the interval $[0,\sqrt{2}]$ with a condition
$R+T \leq \sqrt{2}$.
The conformal diagram representing our analog spacetime is  depicted in Fig.\ \ref{fig3a}. 
The singularity at $\tau=\tau_{\rm c}$ is mapped onto the segment $[0,\sqrt{2}]$
on the horizontal axis.

The coordinate transformation 
\begin{equation}
t'=\int \beta d\tau
\label{eq138}
\end{equation}
brings the metric (\ref{eq018})
to the standard form of an open $k=-1$ FRW  spacetime metric
with the cosmological time 
$t'$. The time coordinate $t'$ is related to the original time $t$ via $\tau$ and the transformation
(\ref{eq147}).
The analog cosmological scale is  $a(t')=\alpha(\tau(t'))/r_0$, where
the constant $r_0$ is related to the spatial Gaussian curvature 
$K=-1/r_0^2$.
The proper distance is   $d_{\rm p}=\alpha y$ and the analog Hubble constant is
\begin{equation}
{\cal H}=\frac{\dot{\alpha}}{\alpha\beta},
\label{eq238}
\end{equation}
where the overdot denotes a partial derivative with respect to $\tau$. 
Then, we define 
the {\em analog Hubble horizon}  as a two-dimensional spherical surface at which the 
magnitude of the analog recession velocity
\begin{equation}
v_{\rm rec}= {\cal H}d_{\rm p}  = y\frac{\dot{\alpha}}{\beta}
\label{eq038}
\end{equation}
equals the velocity of light. Hence, 
the condition
\begin{equation}
y =\frac{\beta}{\left|\dot{\alpha}\right|}
\label{eq318}
\end{equation}
defines the location of the analog Hubble horizon.
Note that the radial fluid velocity $v$ in (\ref{eq246}) and
the analog recession velocity (\ref{eq038}) are quite distinct quantities --
in an expanding fluid  $v$ is  always positive and less than 
the velocity of light $c=1$, whereas $v_{\rm rec}$ may be positive or negative
depending on the sign of $\dot{\alpha}$ and its
 magnitude  may be arbitrarily large.

A two-dimensional spherical
 surface  on which the radial  velocity $v$ equals the velocity of pions $c_\pi$
defines another horizon, which we refer to
as the {\em naive Hubble horizon}.
This horizon is obviously distinct from the  analog Hubble horizon defined above.
The evolution of the naive and the analog Hubble horizons  with $\tau$ are
depicted in  Fig.~\ref{fig3}.

Next we  introduce the concept of an
analog mar\-gi\-nal\-ly trapped surface or 
{\em analog apparent horizon}  following closely the general considerations of Sec.~\ref{radial}
and the Appendix \ref{trapped}.
The formation of an analog apparent horizon in an expanding hadronic fluid 
is similar to the formation of a black hole in a gravitational collapse
although the role of  an outer trapped surface is exchanged with that of 
an inner trapped surface.
Unlike a black hole in general relativity, the formation of which is indicated by the existence
of an   {\em outer} marginally trapped surface,
the formation of an analog black (or white) hole in an expanding fluid 
is indicated by the existence of a  future  or past {\em inner} marginally  trapped surface.

Equation (\ref{eq117}) with (\ref{eq108}) defines a  hypersurface which we refer to as the 
{\em analog trapping horizon}.
Any solution to Eq.\ (\ref{eq117}) , e.g.,  in terms of $r$ for fixed $t$, gives the location of 
the analog apparent horizon $r_H$. For example, the radius $r_H=1.53\tau_0$  computed using (\ref{eq117}) 
for fixed $t=6t_0$ is the point of vanishing outgoing null expansion which marks the location of the apparent horizon
in the top left panel of Fig.\ \ref{fig2}.
From (\ref{eq110}) it follows that the region of spacetime
 for which 
 \begin{equation}
\tanh y \geq |\beta/\dot{\alpha}|
\label{eq158}
\end{equation}
is trapped.
It is future trapped 
if $\dot{\alpha} < 0$  and past trapped if $\dot{\alpha} > 0$.
The condition (\ref{eq158}) can be met only if
$|\beta/\dot{\alpha}| \leq 1$, which holds for $\tau$  between
$\tau_{\rm c}$ and  $\tau_{\rm max}$. At 
$\tau=\tau_{\rm max}$ we have $|\beta/\dot{\alpha}|=1$,
so the endpoint of the trapping horizon
 in Fig. \ref{fig3}
is at
$\tau=\tau_{\rm max}$, $\tanh y=1$.

We find that $\dot{\alpha}$ is negative for $\tau$ in 
the entire range $\tau_{\rm c} \leq \tau \leq \tau_{\rm max}$ and,
according to (\ref{eq238}),
the analog Hubble constant is always negative.
Hence, our
analog cosmological model is a contracting FRW spacetime with a negative spatial curvature.
The shaded area  
  left of the bold line 
in Fig.\ \ref{fig3}  represents the time evolution of the future trapped region.
Note that the analog Hubble horizon is always behind the apparent horizon whereas
 the naive Hubble horizon may be located ahead of or behind the apparent horizon
 depending on the magnitude of $\dot\alpha$.
The naive Hubble and  apparent horizons coincide if 
$a$ and $c_\pi$ are $\tau$ independent constants.

The apparent horizon is generally not a Killing horizon  and normally does not coincide with the event horizon
(one exception
is de Sitter spacetime).
Moreover, the apparent horizon exists in all FRW universes \cite{ellis}, 
whereas  the event horizon does not exist  in eternally expanding FRW universes 
with the equation of state $w>-1/3$ (see, e.g., \cite{davis}). 
For the metric (\ref{eq018}), the event horizon 
is defined by
\begin{equation}
y= \int_\tau^\infty d\tau' \frac{c_\pi (\tau')}{\tau'}.
\label{eq207}
\end{equation}
In our chiral fluid model 
the integral on the right-hand side diverges at the upper limit because 
$c_\pi\rightarrow 1$ as  $\tau\rightarrow \infty$ and hence, the analog event horizon does not exist.
In contrast, as we have demonstrated, the analog apparent horizon does exist.

\subsection{Analog Hawking effect}
\label{surface}
One immediate effect related to the  apparent horizon is the 
Hawking radiation.
Unfortunately,
in a nonstationary spacetime,
the surface gravity associated to the apparent horizon is not uniquely defined
\cite{nielsen2}.
Several ideas
have been put forward how to generalize the definition of surface gravity for the
case when the apparent horizon does not coincide with 
the event horizon
\cite{fodor,mukohyama-booth,hayward,hayward2}.
In this paper we use  the  prescription of 
\cite{hayward2} which we have adapted to analog gravity in our previous paper \cite{tolic}.
This prescription involves the so-called Kodama vector $K^\mu$ \cite{kodama} which generalizes 
the concept of the time translation Killing vector to nonstationary spacetimes.
The analog surface gravity $\kappa$ is defined by 
\begin{equation}
\kappa =\frac{1}{2} \frac{1}{\sqrt{-h}} \partial_\alpha ( \sqrt{-h}h^{\alpha\beta}k n_\beta),
\label{eq228}
\end{equation}
where the quantities on the righ-thand side should be evaluated on the trapping horizon.
The  metric $h_{\alpha\beta}$  of the two-dimensional
space normal to the surface of spherical symmetry  and the 
vector $n_\alpha$ normal to that surface are given by (\ref{eq122}) and (\ref{eq114}), respectively.

The definition (\ref{eq228})  differs from the original 
expression for the dynamical surface gravity \cite{hayward2,hayward3}
by   a normalization factor $k$  which we have introduced in order to meet the requirement that $K^\mu$
should coincide with the time translation Killing
vector $\xi^\mu$  for a stationary geometry.  For the metric (\ref{eq018})
with (\ref{eq108}) we have found \cite{tolic}
\begin{equation}
k= \beta \left(\cosh^2y-\sinh^2y \frac{\tau \dot{\alpha}}{\alpha}\right).
\label{eq324}
\end{equation}
Then, the definition (\ref{eq228}) yields 
\begin{equation}
\kappa =
\frac{v}{2 \beta\gamma(\alpha -\tau \dot{\alpha}v^2)^2}
\left[\alpha(\dot{\alpha}^2+\alpha\ddot{\alpha}-\beta^2)
+2\beta^2 \left(\tau\dot{\alpha}-\alpha\right)v
+ (\alpha \dot{\alpha}^2-2\tau \dot{\alpha}^3+\beta^2 \tau \dot{\alpha}) v^2
\right]
\label{eq229}
\end{equation}
evaluated on the trapping horizon.
The above expression  may be somewhat simplified
by making use of the horizon condition
(\ref{eq117}). We find
\begin{equation}
\kappa =
\frac{c_\pi}{2\tau }
\frac{1 + 2c_\pi v (1-v)- (2+c_\pi) v^3}{\gamma v(1+c_\pi v )^2}
+\frac{\ddot{\alpha}}{2\beta}\frac{v}{\gamma (1+c_\pi v)^2} 
\label{eq231}
\end{equation}
evaluated on the trapping horizon.

It is worthwhile analyze the limitting case of (\ref{eq229})
when the quantities  $a$  and  $c_\pi$ are constants.
Then  $\dot{\alpha}=\alpha/\tau$, $\ddot{\alpha}=0$, and  the apparent horizon
is fixed by the condition $v = c_\pi$. At any chosen time $t=\tau (1-c_\pi^2)^{-1/2}$ the horizon 
is located at $r_H=c_\pi t$
and the expression for $\kappa$  reduces to
\begin{equation}
\kappa =\frac{1}{2t}=\frac{\sqrt{1-c_\pi^2}}{2\tau} 
\label{eq230}
\end{equation}
Hence, the analog surface gravity is finite for any physical value of $c_\pi$ and is maximal when $c_\pi=0$.
However, with $c_\pi=0$ the horizon degenerates to a point located at the origin $r=0$.

\begin{figure}[t]
\begin{center}
\includegraphics[width=0.9\textwidth,trim= 0 0cm 0 0cm]{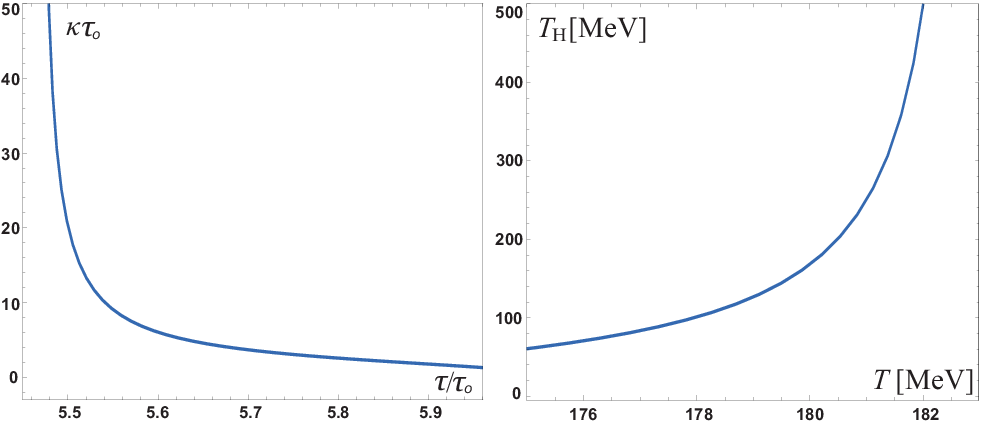}
\caption{Analog surface gravity as a function of the
proper time $\tau$ (left panel)  and the corresponding Hawking temperature  as a function of 
the  fluid temperature $T$ (right panel).}
 \label{fig4}
\end{center}
\end{figure}

In the left panel of Fig.\ \ref{fig4} we plot $\kappa$ as a  function of $\tau$ as given by (\ref{eq231}).
The corresponding temperature defined as
\begin{equation}
T_H=\frac{\kappa}{
 2\pi}
\label{eq044}
\end{equation}
 represents
the analog Hawking temperature of thermal pions emitted at the apparent horizon
as measured by an observer near
 infinity. Since the background geometry is flat, this temperature  equals
 the locally measured Hawking temperature at the horizon.
Thus,  Eq.
(\ref{eq044}) with (\ref{eq229})
corresponds to the flat spacetime
Unruh effect.

As we move along the trapping horizon the radius of the apparent horizon increases
and the Hawking  temperature decreases rapidly with $\tau$.
Hence, there is a correlation between $T_H$ and the local fluid temperature $T$
which is related to $\tau$ by (\ref{eq007}).
In the right panel of Fig.\ \ref{fig4} we show the Hawking temperature $T_H$
as a function of the fluid temperature $T$ at the apparent horizon.

In our previous paper \cite{tolic}  
we showed that
the surface gravity diverges  as 
\begin{equation}
\kappa = (\eta+1/2)(\tau-\tau_{\rm c})^{-1} 
\label{eq048}
\end{equation}
at the singular  point, where $\eta$ is a  constant related to the
scaling of the quantity $\sqrt{a/c_\pi}$
\begin{equation}
\sqrt{\frac{a}{c_\pi}}\propto  (T_{\rm c}-T)^{-\eta}
\label{eq045}
\end{equation}
in the neighborhood of
the critical point.
The constant  $\eta$ may be roughly estimated as follows.
The estimate of the function $\Sigma (q)$ defined in
(\ref{eq203})
 in the neighborhood of
$q^\mu=0$  for small $\sigma$ yields \cite{son2}
\begin{equation}
\Sigma(0, \mbox{\boldmath $q$}^2) \sim 
\frac{T}{\sigma}  \mbox{\boldmath $q$}^2; \hspace{1cm} \Sigma(q_0, 0) \sim \frac{T^2}{\sigma^2} q_0^2
\label{eq248}
\end{equation}
By comparing this with (\ref{eq202}) we deduce the behavior of the  quantities $a$ and $b$ 
for small $\sigma$
\begin{equation}
a \sim 
\frac{T}{\sigma}  ; \hspace{1cm} a+b \sim \frac{T^2}{\sigma^2}.
\label{eq249}
\end{equation}
Then, from (\ref{eq015})
 the pion velocity goes to zero approximately as $c_\pi \propto \sigma^{1/2}$
whereas the ratio
$a/c_\pi$ diverges as $a/c_\pi \propto \sigma^{-3/2}$.
From Eq. (\ref{eq032}) we find  $\sigma \propto (T_{\rm c}-T)^{1/2}$ near the critical point
which yields  $\eta=3/8$.

Numerically, by fitting $\sqrt{a/c_\pi}$ in the close neighborhood of $T_{\rm c}$
to the function
(\ref{eq045})
with the critical temperature $T_{\rm c}=182.822$ MeV obtained numerically from (\ref{eq032}), we find $\eta=0.253$.
A more refined analysis  
based on scaling and universality arguments of Son and Stephanov \cite{son1}
yields  $\eta=0.1975$ \cite{tolic}.

\section{Summary and discussion}
\label{conclusion}
We have demonstrated that,
owing to
the analog gravity effects in
high energy collisions,
a close analogy  may be drawn 
between the evolution of
a hadronic fluid and the spacetime expansion.
Using the formalism of relativistic acoustic geometry we have analyzed the 
expanding chiral fluid in the regime of broken chiral symmetry.
 The expansion which takes place after the collision  
is modeled by  spherically  symmetric Bjorken type expansion.
 The propagation of massless pions in the chiral
fluid provides a geometric  analog of expanding spacetime
equivalent to an open  ($k=-1$)  FRW cosmology.
The geometry depends on the parameters $a$ and $b$ of the effective Lagrangian defined in Sec.\
\ref{chiral}.
The elements of the analog  metric tensor 
are functions of the spacetime coordinates via the temperature dependence of $a$ and 
the pion velocity $c_\pi$.
The pions propagate slower than light with a velocity close to zero in the neighborhood of
the critical point of the chiral phase transition.

A trapped region forms for radial velocities of the fluid beyond  the  value
defined 
 by Eq. (\ref{eq117}). This value  defines a hypersurface 
shown in Fig.\ \ref{fig3} which we refer to as the analog trapping horizon, 
at which the outgoing radial  null expansion vanishes.
Our trapping horizon is foliated by future inner marginally trapped surfaces and is equivalent to
the trapping horizon in
contracting FRW spacetime, i.e., in dynamical spacetime with a negative Hubble constant.
The shaded area in Fig.\ \ref{fig3} represents the time evolution of 
the future trapped region,
with the future inner marginally trapped surface (or the future apparent horizons) as its inner boundary.
This marginally trapped surface  may be regarded as an ``outer'' white hole: the ingoing  pions (future directed ingoing null geodesics)
freely cross the apparent horizon whereas the outgoing cannot penetrate the apparent horizon.
This is opposite to an expanding FRW universe where the inner marginally trapped surface 
 acts as a black hole: the future directed ingoing null geodesics cannot escape the apparent horizon
whereas the outgoing null geodesics freely cross the apparent horizon.

We have studied the Hawking effect associated with the analog  apparent horizon
using the Kodama-Hayward definition of surface gravity adapted to
the analog gravity geometry. The Hawking temperature correlates with the local temperature of the fluid
at the apparent horizon and diverges at the critical point.
In contrast to the usual general relativistic Hawking effect, where the Hawking temperature 
is tiny compared with the temperature of the background, the analog horizon temperature 
is of the order or even larger than the local temperature of the fluid.

The most important outcome of our analysis relevant for particle physics phenomenology is thermal radiation of pions due to the analog Hawking effect.In that regard, it is tempting to speculate about possible signals for the analog Hawking effect in high energy collisions. In principle, one could  measure the temperature by fitting the pion spectrum to the thermal  Planck distribution. However, one additional signal should be invented in order to unambiguously  distinguish between the thermal pions produced above the critical temperature from those emitted as an analog Hawking radiation from the apparent horizon below the critical temperature.

The analog Hawking radiation of pions  should not be confused   with the 
Hawking-Unruh radiation of hadrons of Castorina {\em et al.}~\cite{castorina}.
The latter is a usual Unruh effect due to the acceleration of  quark-antiquark pairs
produced in particle collisions, whereas the former is an analog thermal radiation due to effective geometry of the chiral fluid.

A spherically symmetric Bjorken expansion model considered here 
may be phenomenologically viable as a model of hadron production in $e^+e^-$,
but it is certainly not the best model for description 
 of high energy heavy ion collisions.
It would be desirable to apply our formalism to a more realistic hydrodynamic model
that involves 
a transverse expansion
superimposed on a longitudinal boost invariant expansion.
In this case the calculations become rather involved, as the formalism for general nonspherical
spacetimes is not yet fully developed. This work is in progress.

In conclusion, we believe that the study of analog gravity in high energy
collision may in general improve our understanding of both particle physics phenomenology
and dynamical general relativistic systems.

\subsection*{Acknowledgments}
This work was supported by the Ministry of Science,
Education and Sport
of the Republic of Croatia under Contract No. 098-0982930-2864
and 
partially supported by the ICTP-SEENET-MTP Grant No. PRJ-09 ``Strings and Cosmology`` 
in the frame of the SEENET-MTP Network.

\appendix

\section{Trapped surfaces in general relativity}
\label{trapped}

Following Hayward, \cite{hayward} we summarize here  the  relevant definitions
related to trapped surfaces.

\begin{enumerate}
\item {\em Trapped surface}:
Let $\Sigma$ denote a spacelike hypersurface, e.g., a hypersurface defined by $t=\rm const$.
A two-dimensional surface $S\subset \Sigma$ with spherical topology is called a {\em trapped surface} on $\Sigma$ if 
the families of ingoing and outgoing null geodesics normal to the surface are  both converging or both diverging.
More precisely, 
 the null expansions
$\varepsilon_+=l^\mu_{+;\mu}$ and $\varepsilon_-=l^\mu_{-;\mu}$ on a trapped surface $S$ should satisfy
$\varepsilon_+\varepsilon_- >0$.
One  distinguishes between 
a {\em past trapped surface}  for which  
both $\varepsilon_+$ and $\varepsilon_-$ are positive, and a {\em future trapped surface} 
for which  both $\varepsilon_+$ and $\varepsilon_-$ are negative. 
\item
{\em Trapped region}:
A set of future  trapped surfaces (or closed trapped surfaces \cite{ellis,hawking}) on $\Sigma$ is
referred to as  a {\em future   trapped region}. Similarly, a set of past  trapped surfaces  on $\Sigma$ is
called a  {\em past trapped region}.
\item
{\em Marginally trapped surface}:
A two-dimensional surface $H$ is said to be {\em marginally trapped} if one of the null expansions vanishes on $H$,
i.e., if either  $\varepsilon_+|_H=0$ or $\varepsilon_-|_H=0$. 
A marginally trapped surface is also referred to as an {\em apparent horizon} 
although, strictly speaking, the original definition of Ellis and Hawking \cite{hawking}
involves in addition the assumption of  asymptotic flatness.

\item
{\em  Outer marginally trapped surface}:
 A surface $H$ is said to be future (past) outer marginally trapped 
 if  on $H$ 
the future (past) directed  outgoing  null expansion  
vanishes, its derivative along the ingoing null geodesic is negative, and the ingoing 
null expansion is negative, i.e.,
if  $\varepsilon_+|_H=0$,  $l_-^\mu\partial_\mu\varepsilon_+|_H <0$, and $\varepsilon_-|_H<0$.
Equivalently, the future (past) outer marginally trapped surface may be defined as the surface on which
the past (future) directed  ingoing  null expansion  
vanishes, its derivative along the outgoing null geodesic is negative, and the outgoing 
null expansion is positive, i.e.,
if $\varepsilon_-|_H=0$, $l_+^\mu\partial_\mu\varepsilon_-|_H <0$, and $\varepsilon_+|_H>0$.

\item
{\em  Inner marginally trapped surface}:
A surface $H$ is said to be future (past) inner marginally trapped 
if on $H$ the future (past) directed outgoing null expansion  
vanishes, its derivative along the ingoing null geodesic is positive, and the ingoing 
null expansion is negative, i.e.,
if  $\varepsilon_+|_H=0$, $l_-^\mu\partial_\mu\varepsilon_+|_H>0$, and $\varepsilon_-|_H<0$.
Equivalently, the future (past) inner marginally trapped surface may be defined as the surface on which
the past (future) directed  ingoing  null expansion  
vanishes, its derivative along the outgoing null geodesic is positive, and the outgoing 
null expansion is positive, i.e.,
($\varepsilon_-|_H=0$, $l_+^\mu\partial_\mu\varepsilon_-|_H >0$, and $\varepsilon_+|_H>0$).
\item
{\em Trapping horizon}:
A hypersurface foliated by inner or outer marginally trapped surfaces is referred to as
an inner or outer {\em trapping horizon}, respectively.
\end{enumerate}

According to this classification we distinguish four physically relevant classes:
\renewcommand{\theenumi}{\roman{enumi}}%
\begin{enumerate}
 \item 
{\em A future outer marginally trapped surface}
is the outer boundary of a future trapped region 
and is typical of a black hole.
\item 
{\em A past outer marginally  trapped surface} is the outer boundary of a past trapped region
and is typical of a white hole.
\item 
{\em A future inner marginally trapped surface} is the inner boundary of a future trapped region
 and  represents an ''outer'' white hole. This situation 
is physically relevant in the cosmological context for a shrinking universe, i.e., for an FRW spacetime with
$\dot{a} < 0$.
\item 
{\em A past inner marginally trapped surface} is the inner boundary of a past trapped region
and  represents an ``outer'' black hole. This situation
is physically relevant in the context of an expanding FRW universe \cite{ellis,faraoni}.
\end{enumerate}


\begin{thebibliography}{99}
\bibitem{heinz}
  U.~W.~Heinz, C.~Shen, and H.~Song,
   in the AIP Conference Proceedings PANIC11 MIT, 24-29 July 2011 (to be published).
\bibitem{floris}
  M.~Floris (ALICE collaboration),
  J.\ Phys.\ G G {\bf 38}, 124025 (2011); arXiv:1108.3257.
\bibitem{cern}
 CERN Courier, {\bf 52}, January/February 2012, p. 8
\bibitem{shen}
  C.~Shen, U.~Heinz, P.~Huovinen, and H.~Song,
  Phys.\ Rev.\ C {\bf 84}, 044903 (2011).
\bibitem{dumitru-kolb}
A.~Dumitru, 
  Phys.\ Lett.\ B {\bf 463}, 138 (1999);
P.~F.~Kolb, U.~W.~Heinz, P.~Huovinen, K.~J.~Eskola, and K.~Tuominen,
  Nucl.\ Phys.\ A {\bf 696}, 197 (2001).
\bibitem{bjorken} 
  J.~D.~Bjorken,
  Phys.\ Rev.\ D {\bf 27}, 140 (1983).
\bibitem{tuts}
M.~Tuts, CNN, 9 November 2010.
\bibitem{bbc}
BBC News, 8 November 2010.
\bibitem{evans1}
D.~Evans,  The Telegraph, 8 November 2010.
\bibitem{barcelo}
  C.~Barcelo, S.~Liberati, and M.~Visser,
  Living Rev.\ Rel.\  {\bf 14}, 3 (2011); gr-qc/0505065.
\bibitem{visser}
M.~Visser,
 Class. Quant. Grav. {\bf 15}, 1765--1791 (1998).
%
\bibitem{philbin}
  T.~G.~Philbin, C.~Kuklewicz, S.~Robertson, S.~Hill, F.~Konig, and U.~Leonhardt,
  Science {\bf 319}, 1367 (2008).
\bibitem{jacobson}
T.A.~Jacobson and G.E.~Volovik,
Phys.\ Rev.\ D {\bf 58}, 064021 (1998).
\bibitem{moncrief}
V.\ Moncrief, 
      Astrophys.\ J. {\bf 235}, 1038-46 (1980).
\bibitem{abraham} 
  H.~Abraham, N.~Bili\'c and T.~K.~Das,
  Class.\ Quant.\ Grav.\  {\bf 23}, 2371 (2006); gr-qc/0509057; 
  T.~K.~Das, N.~Bili\'c and S.~Dasgupta,
 JCAP {\bf 0706}, 009 (2007); astro-ph/0604477.
\bibitem{tolic} 
  N.~Bili\'c and D.~Toli\'c,
 Phys.\ Lett.\ B {\bf 718}, 1 (2012), arXiv:1207.2869,
\bibitem{shifman} 
M.A.~Shifman,   Ann.\ Rev.\ Nucl.\ Part.\ Sci.\ {\bf 33}, 199 (1983).   
\bibitem{harris}
    J. W. Harris and B. M\"uller,
    Ann.\ Rev.\ Nucl.\ Part.\ Sci.\ {\bf 46}, 71 (1996).
\bibitem{bilic}
    N.\ Bili\'c, J.\ Cleymans, and M.\ D.\ Scadron,
    Int.\ J.\ Mod.\ Phys.\ {\bf A10}, 1169 (1995).
%
\bibitem{bilic1}
    N.\ Bili\'c and H.\ Nikoli\'c,
    Eur.\ Phys.\ J.\ C {\bf 6}, 515 (1999).
%
\bibitem{gell}
    M.\ Gell-Mann and M.\ L\'{e}vy,
    Nuovo Cimento {\bf 16}, 705 (1960).

\bibitem{pisarski2}
    R.\ D.\ Pisarski and M.\ Tytgat,
    Phys.\ Rev.\ D {\bf 54}, R2989 (1996).
%
\bibitem{son1}
    D.\ T.\ Son  and M.\ A.\ Stephanov,
    Phys.\ Rev.\ Lett.\ {\bf 88}, 202302 (2002).
%
\bibitem{son2}
    D.\ T.\ Son  and M.\ A.\ Stephanov,
    Phys.\ Rev.\ D {\bf 66}, 076011 (2002).
\bibitem{lampert} 
  M.~A.~Lampert, J.~F.~Dawson and F.~Cooper,
  Phys.\ Rev.\ D {\bf 54}, 2213 (1996);
  G.~Amelino-Camelia, J.~D.~Bjorken and S.~E.~Larsson,
  Phys.\ Rev.\ D {\bf 56}, 6942 (1997);
  M.~A.~Lampert and C.~Molina-Paris,
  Phys.\ Rev.\ D {\bf 57}, 83 (1998);
  A.~Krzywicki and J.~Serreau,
   Phys.\ Lett.\ B {\bf 448}, 257 (1999).
\bibitem{bilic2} 
  N.~Bili\'c and H.~Nikoli\'c,
  Phys.\ Rev.\ D {\bf 68}, 085008 (2003); hep-ph/0301275.
%
\bibitem{meyer}
    H.\ Meyer-Ortmanns and B.-J. Schaefer,
    Phys.\ Rev.\ D {\bf 53}, 6586 (1996). 
\bibitem{rod}
    D.\ R\"oder, J.\ Ruppert, and D.\ H.\ Rischke,
    Phys.\ Rev.\ D {\bf 68}, 016003 (2003).
%
%
\bibitem{pis}
    R.\ D.\ Pisarski and F.\ Wilczek,
    Phys.\ Rev.\ D {\bf 29}, 338 (1984).
\bibitem{baacke} 
  J.~Baacke and S.~Michalski,
  Phys.\ Rev.\ D {\bf 67}, 085006 (2003).
\bibitem{landau}
 L.D. Landau and E.M. Lifshitz,
 {\em Statistical Physics},
(Pergamon, Oxford, 1993) p. 187.
\bibitem{kolb-russkikh} 
  P.~F.~Kolb, J.~Sollfrank, and U.~W.~Heinz,
  Phys.\ Rev.\ C {\bf 62}, 054909 (2000).
  V.~N.~Russkikh and Y.~.B.~Ivanov,
  Phys.\ Rev.\ C {\bf 76}, 054907 (2007).
%
%

\bibitem{bilic3}
N.\ Bili\'c, Class.\ Quantum Grav.\ {\bf 16}, 3953 (1999). 
\bibitem{visser2}
M.~Visser and C.~Molina-Paris, New J.\ Phys.\ {\bf 12}, 095014 (2010).  
%
\bibitem{ellis}
  G.~F.~R.~Ellis,
  Gen.\ Rel.\ Grav.\  {\bf 35}, 1309 (2003).
\bibitem{davis} 
  T.~M.~Davis and C.~H.~Lineweaver,
PASA, 21, 97-109 (2004); astro-ph/0310808.
  \bibitem{nielsen2} 
  A.~B.~Nielsen and J.~H.~Yoon,
  Class.\ Quant.\ Grav.\  {\bf 25}, 085010 (2008).
\bibitem{fodor} 
  G.~Fodor, K.~Nakamura, Y.~Oshiro, and A.~Tomimatsu,
  Phys.\ Rev.\ D {\bf 54}, 3882 (1996).
\bibitem{mukohyama-booth} 
  S.~Mukohyama and S.~A.~Hayward,
  Class.\ Quant.\ Grav.\  {\bf 17}, 2153 (2000);
 I.~Booth and S.~Fairhurst,
 Phys.\ Rev.\ Lett.\  {\bf 92}, 011102 (2004).
\bibitem{hayward}
  S.~A.~Hayward,
  Phys.\ Rev.\ D {\bf 49}, 6467 (1994).
\bibitem{hayward2} 
  S.~A.~Hayward,
  Class.\ Quant.\ Grav.\  {\bf 15}, 3147 (1998).
\bibitem{kodama} 
  H.~Kodama,
  Prog.\ Theor.\ Phys.\  {\bf 63}, 1217 (1980).
\bibitem{hayward3}
  S.~A.~Hayward, R.~Di Criscienzo, L.~Vanzo, M.~Nadalini, and S.~Zerbini,
  Class.\ Quant.\ Grav.\  {\bf 26}, 062001  (2009).
\bibitem{castorina} 
  P.~Castorina, D.~Kharzeev and H.~Satz,
  Eur.\ Phys.\ J.\ C {\bf 52}, 187 (2007).
\bibitem{hawking}
S.W.~Hawking and G.F.R.~Ellis,
{\it The large scale structure of space-time}
(Cambridge University Press,
    Cambridge, 1973).
\bibitem{faraoni} 
 V.~Faraoni,
  Phys.\ Rev.\ D {\bf 84}, 024003 (2011).
  

\end{thebibliography}
\end{document}